\documentclass[pdftex,inlucegraphix,usenatbib]{jetpl}

\twocolumn

\lat

\def\prd{Physical Review D }
\def\jcap{JCAP }

\def\apj{ApJ }

\def\mnras{MNRAS }

\def\sovast{Sov. Astron. }

\def\Mch{{\cal M}}
\newcommand{\mpl}{m_\mathrm{Pl}}

\def\beq#1{\begin{equation}\label{#1}}
\def\eeq{\end{equation}}
\def\beqa#1{\begin{eqnarray}\label{#1}}
\def\eeqa{\end{eqnarray}}

\def\Eq#1{Eq.~(\ref{#1})} 

\def\myfrac#1#2{\left(\frac{#1}{#2}\right)}

\def\comment#1{\relax}

\title{Coalescing primordial binary black holes with lognormal mass spectrum}

\rtitle{PBH with lognormal mass spectrum}

\sodtitle{Coalescing primordial binary black holes with log-normal mass spectrum}

\author{K.\,A.\,Postnov$^{+*}$\/\thanks{e-mail: pk@sai.msu.ru}
A.\,D.\,Dolgov$^*$, 
N.\,A.\, Mitichkin$^+\dag$, 
I.\,V.\, Simkin$^{\ddag*}$}

\rauthor{K.\,A.\,Postnov, A.\,D.\,Dolgov, N.\,A.\, Mitichkin, I.\,V.\, Simkin }

\sodauthor{Postnov, Dolgov, Mitichkin, Simkin}

\address{$^+${Sternberg Astronomical Institute, M.V. Lomonosov Moscow State University,\\ 13, Universitetskij pr., 119234, Moscow, Russia}\\~\\
$^*${Department of Physics, Novosibirsk State University, \\Pirogova 2, 630090, Novosibirsk, Russia}\\~\\
$^\dag${Faculty of physics, M.V.Lomonosov Moscow State University, Leninskie Gory 1, Moscow, Russia}\\~\\
$^\ddag${Bauman Moscow State Technical University, Moscow, Russia}
}

\dates{January 2021}{January 2021}

\abstract{Primordial black holes created in the early Universe can constitute a substantial fraction of dark matter and serve as seeds for early galaxy formation. Binary primordial black holes with masses of the order of a few dozen solar masses can explain the observed LIGO/Virgo gravitational-wave events. In this Letter, we show that primordial black holes with log-normal mass spectrum centered at $M_0\simeq 15-17 M_\odot$ simultaneously explain  both the chirp mass distribution of the detected LIGO/Virgo binary black holes and the differential chirp mass distribution of merging binaries as inferred from the LIGO/Virgo observations. The obtained parameters of log-normal mass spectrum of primordial black holes also give the fraction of seeds with $M\gtrsim 10^4 M_\odot$ required to explain the observed population of supermassive black holes at $z=6-7$.}


\begin{document}

\maketitle 

Primordial black holes (PBHs) were first introduced by Zeldovich and Novikov \cite{1967SvA....10..602Z} and Carr and Hawking \cite{1974MNRAS.168..399C}. They considered energy density fluctuations $\delta \rho /\rho$ at the radiation-dominated (RD) stage whose amplitude at the cosmological horizon can be unity. In this case, the mass
of the fluctuation is such that its radius is smaller than the gravitational radius $R\le R_g=2GM/c^2$ (here $G$ and $c$ are the gravitational constant and speed of light, respectively), i.e. such a fluctuation gives rise to a black hole. Below we will use the natural units $G=c=\hbar=k_B=1$, where  $\hbar$ and $k_B$ are the Dirac and Boltzmann constants, respectively. In these units, $G=1/\mpl^2$, where $\mpl = 2.2 \times 10^{-5}$~g $= 2 \times 10^{43}$~s$^{-1}$ ={ $1.22\times 10^{19}$~GeV} is the Planck mass. 

Therefore, the conventional condition for PBH formation from an accidentally large primordial energy density fluctuation can be written as follows. At RD stage, the mass comprised inside the cosmological horizon $r_h=2t$ is 
\beq{e:Mh}
M_h=\mpl^2t=2.2\times 10^5 M_\odot \myfrac{t}{1\,\mathrm{s}}
\eeq 
This formula shows that depending on the formation time, PBH masses can fall into a very wide range which presently can span from $\sim 10^{15}$~g to billions solar masses. In the present paper, of special interest for us are stellar-mass black holes that can be formed at QCD phase transition stage. Indeed, the temperature at the RD stage is related to the energy density $\rho$ as
\beq{e:rho}
\rho = \frac{3 \mpl^2}{32 \pi t^2} = \frac{\pi^2 g_*}{30} \,T^4\,.
\eeq
\noindent Here $g_* \approx 40$ is the number of relativistic species in the primordial plasma above and close to $T_\mathrm{QCD}$.
Therefore, the mass inside the horizon \Eq{e:Mh} can be expressed as \cite{2020JCAP...07..063D}
\beq{e:Mh8}
M_h\approx 8 M_\odot\myfrac{100\,\mathrm{MeV}}{T_\mathrm{QCD}}^2\myfrac{40}{g_*}^{1/2}
\eeq
\noindent 
where $g_* \approx 40$ is the number of relativistic species in the primordial plasma above and close to $T_\mathrm{QCD}$.

Using (\ref{e:Mh}) and (\ref{e:rho}), the PBH creation condition $R\le 2M/\mpl^2$ for the overdensity $\delta\rho$ can be recast into the form
\beq{e:dh}
\frac{\delta\rho}{\rho}\ge \myfrac{M_h}{M}^2
\eeq
 
Clearly, this is a rather crude criterion for PBH formation from primordial fluctuations. More detailed calculations for the patches of the horizon collapsing into PBH can be performed by assuming a specific shape and type of the fluctuations
(see, e.g., \cite{1975ApJ...201....1C,2013CQGra..30n5009M,2013PhRvD..88h4051H,2019JCAP...12..029K}). 

In \cite{1993PhRvD..47.4244D}, the cosmological inflation has been for the first time employed for PBH formation. It allowed to create PBHs with 
masses exceeding millions solar masses. PBH creation from adiabatic quantum fluctuations arising at the inflation stage was later considered in \cite{1994PhRvD..50.7173I}.
It was based on Starobinsky's suggestion \cite{1992JETPL..55..489S} on the perturbation generation of the inflaton field in a model with double field inflation. The PBH mass spectrum calculated in ref. \cite{1994PhRvD..50.7173I} has a rather complicated analytical structure and to the best of our knowledge has not been applied to the analysis of observational data. 

Depending on the model, PBHs formed at a given epoch can have different mass spectra, including power-law $dn/dM\sim M^{-\alpha}$ \cite{1975ApJ...201....1C}, log-normal $dn/dM\sim \exp[-\gamma\ln(M/M_0)^2]$
\cite{1993PhRvD..47.4244D}, or multipeak \cite{2009NuPhB.807..229D,2019PhRvD..99j3535C} (see recent review \cite{2020arXiv200602838C} for more detail and references). 

In the present paper, we will use the log-normal PBH mass spectrum \cite{1993PhRvD..47.4244D,2009NuPhB.807..229D} that has already been applied to LIGO/Virgo gravitational-wave (GW) events \cite{2016JCAP...11..036B}  and enabled explanation of the LIGO/Virgo binary black hole chirp mass distribution   \cite{2020JCAP...12..017D}. Remind that the chirp mass $\Mch$ is the principal parameter that determines the GW waveform of a coalescing binary system with point-like masses $m_1$ and $m_2$ at the inspiralling phase, $\Mch=(m_1m_2)^{3/5}/(m_1+m_2)^{1/5}$. This parameter is derived from GW observations most accurately. 

\textbf{Cumulative chirp mass distribution of coalescing PBH in LIGO/Virgo observations.} 
As shown in \cite{2020JCAP...12..017D}, the cumulative probability $F(>\Mch)$ to observe a population of binary coalescences at a GW detector with given sensitivity is independent of poorly known fraction of PBH as dark matter $f_\mathrm{PBH}\le 1$. 
At the time of writing of ref. \cite{2020JCAP...12..017D}, only data on chirp masses of sources from O1-O2 LIGO/Virgo runs were available \cite{2019PhRvX...9c1040A}. Therefore, we have estimated chirp masses of O3 sources from open LIGO/Virgo data by employing the fact the actual sensitivity of LIGO/Virgo detectors is primarily determined by the chirp mass of the coalescing binary and defines the limiting luminosity distance to the source (the detector horizon) $D_h\propto \Mch^{5/6}$.

\begin{figure*}
 \begin{center}
\includegraphics[width=0.9\textwidth]{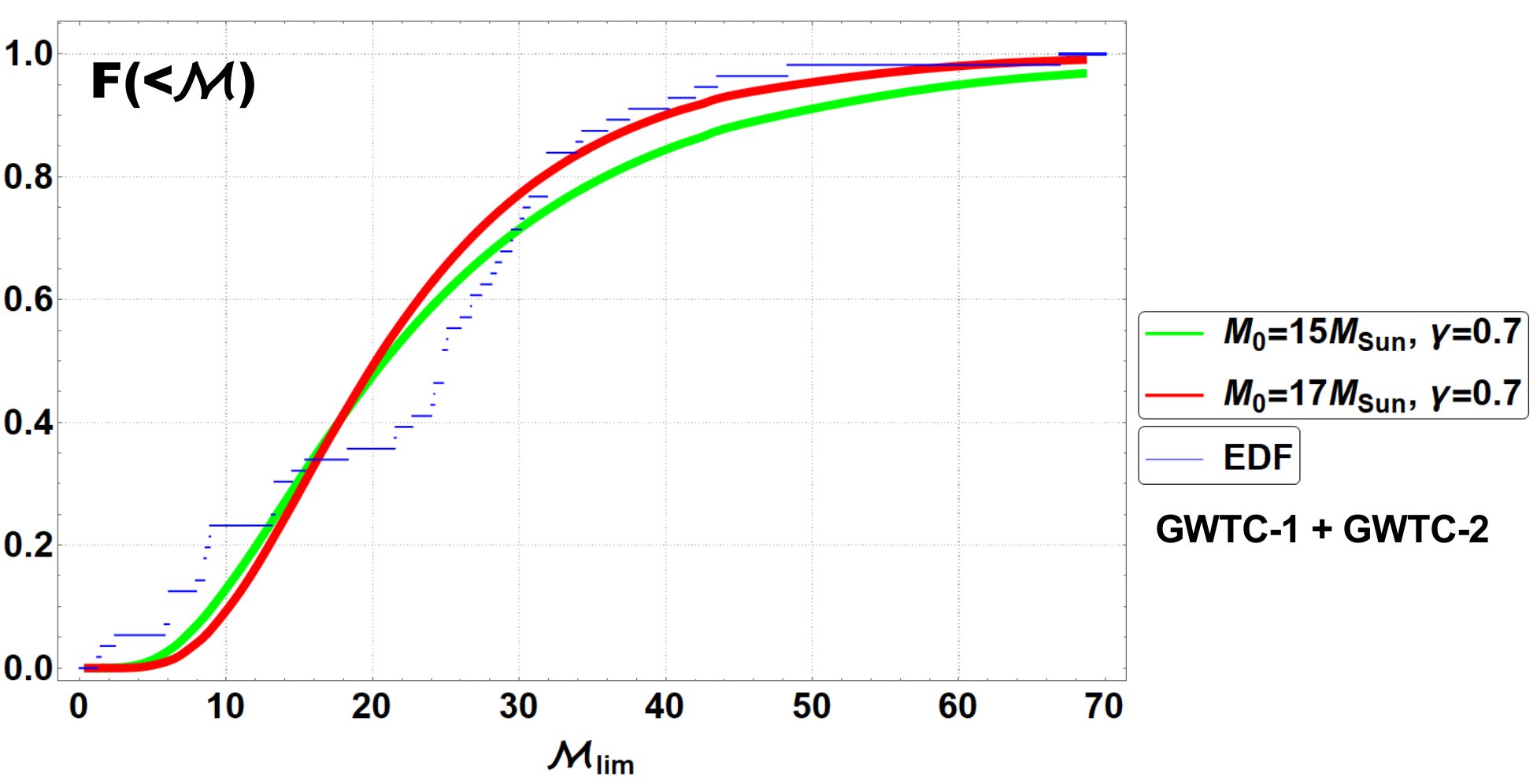}
 \end{center}
  \caption{Fig. 1. Cumulative chirp-mass distribution from LIGO/Virgo GWTC-1 and GWTC-2 catalogs (blue histogram) and predicted 
  chirp-mass distribution from binary PBH coalescences with log-normal PBH mass spectrum (solid curves). The red and green curves show the best-fit PBH mass spectrum parameters $M_0$ and $\gamma$ obtained from Kolmogorov-Smirnov and Van den Waerden statistical tests, respectively, at the 95\% C.L.}
 \label{f:FM}
\end{figure*}

Here we repeat the analysis of ref. \cite{2020JCAP...12..017D} using the published data from both GWTC-1 \cite{2019PhRvX...9c1040A} and GWTC-2 \cite{2020arXiv201014527A} catalogs. The result is presented in Fig. \ref{f:FM}. Notably, the best-fit PBH mass spectrum parameters, $M_0\sim 15-17 M_\odot$ and $\gamma\sim 1$, as inferred from the published GW observations, turned out to be similar to those found in our previous analysis  \cite{2020JCAP...12..017D}. This proves the correctness of our approach to estimate chirp masses from low-latency luminosity distance to the source only. 

\begin{figure*}
 \begin{center}
\includegraphics[width=0.9\textwidth]{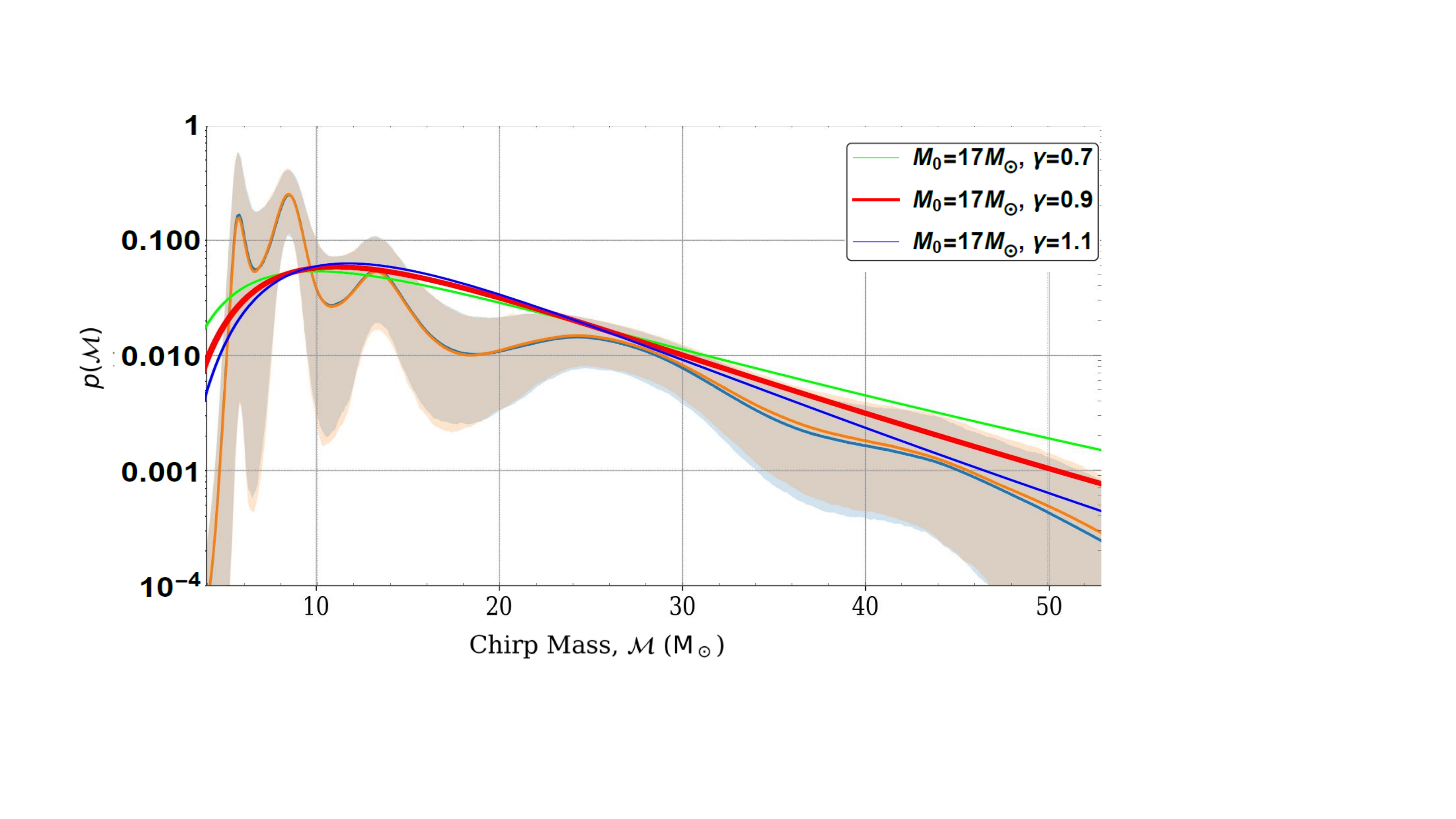}
 \end{center}
  \caption{Fig. 2. The chirp-mass distribution as inferred from LIGO/Virgo observations in ref. \cite{2020arXiv201104502T,2020arXiv201208839T} at 90\% C.L. (grey shaded band around curved orange and blue line) and the expected chirp mass distribution of a merging binary PBH with log-normal mass spectrum of the components. Green, red and blue solid lines correspond to $\gamma=0.7, 0.9, 1.1$, respectively, for $M_0=17 M_\odot$.}
 \label{f:dM}
\end{figure*}

\textbf{Chirp mass distribution in the rest frame of coalescing binaries.}
The GW LIGO/Virgo data enables a Bayesian inference of the intrinsic properties of merging binary BH population. Such an analysis was performed by the LIGO/Virgo team \cite{2020arXiv201014533T} and independently by other authors \cite{2020PhRvD.102l3022R}. Most interesting seems to be the result of ref.  \cite{2020arXiv201104502T,2020arXiv201208839T}, 
which may suggest the emerging evidence for different types of BH populations appeared as BH+BH mergings in distinct mass ranges.
Here we draw attention to the high-mass tail of the chirp mass distribution inferred by Bayesian analysis from the LIGO/Virgo data \cite{2020arXiv201104502T,2020arXiv201208839T}. If the PBH model shown in Fig. \ref{f:FM} is correct, then we would expect that the chirp mass distribution of the merging binaries would be constructed from log-normal mass spectrum of each component. The result is presented in Fig. \ref{f:dM}. 
The grey band in Fig. 2 is taken from \cite{2020arXiv201104502T,2020arXiv201208839T} and shows the Bayesian inference (90\% C.L.) of the chirp masses from the published LIGO/Virgo data assuming a power-law mass distribution. The solid green, red and blue curves present the chirp mass distribution for log-normal mass spectrum of the components with $M_0=17 M_\odot$
 and $\gamma=0.7, 0.9$  and 1.1, respectively. Clearly, the high-mass tail of the inferred chirp mass distribution can be described by log-normal mass spectrum of the components with parameters close to PBH log-normal mass spectrum shown in Fig. 1 and discussed in detail in ref. \cite{2020JCAP...12..017D}. We stress that Fig. 1 and Fig. 2 are totally independent. While Fig. 1 compares the predicted cumulative PBH binary merging chirp mass distribution as observed by the LIGO/Virgo detectors, Fig. 2 shows the the chirp mass distribution in the rest-frame of the coalescing binaries. The high-mass tail in Fig. 2 for the best-fit PBH parameters from Fig. 1 
 (the green curve) lies somewhat higher than the inferred values, but slightly increasing $\gamma$ (red and blue curves) easily 'puts the tail down'. Note also that here we do not address the problem of the Bayesian inference of parameters of the assumed log-normal mass spectrum of BH-BH components from the published LIGO/Virgo observations, which is a separate task.

\textbf{High-mass PBHs as seeds of supermassive BH formation}. Finally, the obtained PBH mass spectrum parameters $M_0$ and $\gamma$
make it possible to estimate the expected fraction of very heavy PBHs that could be seeds of SMBH observed at $z=6-7$. This problem has been addressed in our previous paper \cite{2016JCAP...11..036B}, and here we reappraise this fraction for $M_0=17 M_\odot$ and $\gamma=0.7$. As discussed in ref. \cite{2021arXiv210100209Z}, the number density of luminous quasars with $M\sim 10^9 M_\odot $ is $n<10^{-8}$ Mpc$^{-3}$. The total mass density (both dark matter and baryonic) is $\rho_m\sim 4\times 10^{10} M_\odot$ Mpc$^{-3}$. This means that if PBHs with log-normal mass spectrum constitute a substantial part of dark matter, the fraction of 
PBH with masses larger than $\sim 10^4 M_\odot$, that can be seeds for SMBH growth up to $10^9 M_\odot$ at high redshifts should be at least $10^{-9}$. It is easy to check that for $M_0=15-17 M_\odot$ and $\gamma\sim 1$ this constraint is easily satisfied.

We conclude that PBHs with log-normal mass spectrum predicted in \cite{1993PhRvD..47.4244D} can provide physical explanation to the population properties of binary BH mergings observed by LIGO/Virgo (the chirp-mass distribution at the detector and in the rest-frame of the coalescing binaries). The central mass of the log-normal PBH mass spectrum is found to be around 15-17 solar masses, close to the predicted PBH mass formed at the QCD phase transition from isocurvature fluctuations suggested by the model  \cite{1993PhRvD..47.4244D}.

The authors thank the participants of the INR Theory Department New Year-2021 Seminar for constructive criticism and useful discussions. Partial  
support from RSF grants 19-42-02004 (KP, IS) and 20-42-09010 (AD) is acknowledged. The work of NM is supported by the BASIS foundation.  


\end{document}